\documentclass[twocolumn,showpacs,floats,floatfix,superscriptaddress,aps,pra,nofootinbib]{revtex4}
\usepackage{amsfonts}
\usepackage{amssymb}
\usepackage{amsmath}
\usepackage{bm}
\usepackage{color}
\usepackage{epsfig}
\usepackage{ifthen}
\usepackage{graphicx}
\usepackage{multirow}
\usepackage{bigstrut}
\usepackage{rotating}
\usepackage{longtable}

\def\be{\begin{equation}}
\def\ee{\end{equation}}
\def\bea{\begin{eqnarray}}
\def\eea{\end{eqnarray}}
\def\bse{\begin{subequations}}
\def\ese{\end{subequations}}

\def\ii{i}
\def\marked{s}

\def\ket#1{\vert #1 \rangle}
\def\bra#1{\langle #1 \vert}

\def\M#1#2{M_{\ket{#1}}(#2)}
\def\M#1#2{\mathbf{R}_{#1}(#2)}
\def\sech{\text{sech\,}}

\def\qunit{qudit}
\def\qunits{\qunit s}
\def\Qunit{Qudit}
\def\Qunits{\Qunit s}
\def\qudit{\qunit}

\def\d{d}
\def\qsize{d}
\def\ndata{\mathcal{N}}
\def\ndata{N}
\def\nqubits{n}
\def\ng{N_G}
\def\aver{a}
\def\ini{\aver}
\def\H{\mathbf{H}}
\def\G{\mathbf{G}}
\def\F{\mathbf{F}}
\def\rms{\textsc{rms}}
\def\dark{u}

\def\operator{reflection-about-average}
\def\Operator{Reflection-about-average}

\begin{document}

\author{S. S. Ivanov}
\affiliation{Department of Physics, Sofia University, 5 James Bourchier Boulevard, 1164 Sofia, Bulgaria}
\affiliation{School of Physics and Astronomy, University of St. Andrews, North Haugh, St. Andrews, Fife, KY16 9SS, Scotland}

\author{H. S. Tonchev}
\affiliation{Department of Physics, Sofia University, 5 James Bourchier Boulevard, 1164 Sofia, Bulgaria}

\author{N. V. Vitanov}
\affiliation{Department of Physics, Sofia University, 5 James Bourchier Boulevard, 1164 Sofia, Bulgaria}

\title{Time-efficient implementation of quantum search with qudits}

\begin{abstract}
We propose a simpler and more efficient scheme for the implementation of the multi-valued Grover's quantum search. The multi-valued search generalizes the original Grover's search by replacing qubits with \qunits\ --- quantum systems of $\d$ discrete states.
The \qunit\ database is exponentially larger than the qubit database and thus it requires fewer particles to control.
The Hadamard gate, which is the key elementary gate in the qubit implementation of Grover's search, is replaced by a $\d$-dimensional (complex-valued) unitary matrix $\F$, the only condition for which is to have a column of equal moduli elements irrespective of their phases; it can be realized through any physical interaction, which achieves an equal-weight superposition state.
An example of such a transformation is the $\d$-dimensional discrete Fourier transform, used in earlier proposals; however, its construction is much more costly than that of the far simpler matrix $\F$.
We present examples of how such a transform $\F$ can be constructed in realistic \qunit~systems in a \emph{single} interaction step.
\end{abstract}

\date{\today}


\maketitle

%%%%%%%%%%%%%%%%%%%%%%%%%%%%%%%%%%%%%%%%%%%%%%%%%%%%%%%%%%%%%%%%%%%%%%%%%%%%%%%%%%%%%%%%%%%%%%%%%%%%%%%%%%%%%%%%%%%%%%%%%%
\section{Introduction}

Grover's quantum search algorithm, invented some 15 years ago \cite{Grover}, has become one of the most impressive showcases of quantum computation.
Its efficiency and relative simplicity have made it not only a textbook example of the superiority of quantum computers over their classical counterparts
 but also a promising candidate for a subroutine in various computationally hard problems.
The Grover algorithm finds a marked item in an unstructured database of $\ndata$ items in about $\ng=(\pi/4)\sqrt{\ndata}$ tries with a quantum computer,
 quadratically faster than the classical routine, which requires $O(\ndata)$ tries.
Grover's algorithm can also be adapted to computationally hard problems with structure, by nested quantum searches \cite{Cerf}.
Proof-of-principle Grover search has been demonstrated in nuclear magnetic resonance with two \cite{NMR-4} and three \cite{NMR-8} qubits (corresponding to 4 and 8 database elements),
 in linear optics with 4 elements \cite{Walther}, and in trapped-ion systems with 4 elements \cite{Brickman}.
Because the physical mechanism of Grover's search is amplitude amplification due to constructive wave interference \cite{NC},
 this algorithm has been demonstrated also in individual Rydberg atoms with 8 different levels serving as the database elements \cite{AWB} and in classical Fourier optics with 32 elements \cite{Bhattacharya}.
Although these latter approaches have outperformed the qubit implementations in terms of database size, they are not scalable to large databases.

The quantum computers, if ever built, are expected to outperform the classical computers for \emph{large} databases, with the benefits increasing with the database size $\ndata$.
The dominant model of quantum computing uses qubits --- two-state quantum systems --- linked in quantum circuits with various one- and two-qubit gates.
The size of the Hilbert space for an ensemble of $\nqubits$ qubits is $2^\nqubits$.
The Hilbert space can be increased either by increasing the number of qubits $\nqubits$, or by increasing the number of logical states in each carrier of information and use \textit{\qunits}~instead of qubits \cite{Muthukrishnan}.
There are often practical limitations for increasing the number of qubits; then the use of \qunits~and the ensuing multi-valued quantum logic is a valuable alternative.

\Qunits --- quantum systems with $\d$ states $\ket{q}_0$, $\ket{q}_1$, \ldots, $\ket{q}_{\d-1}$ --- offer exponentially higher dimensionality than qubits.
For example, it has been shown that qutrits --- three-state quantum systems --- provide the best Hilbert-space dimensionality \cite{Greentree}, i.e. database size \emph{vs} ease of control.
It has been shown recently \cite{Vitanov-qutrit} that for some qutrits the most general SU(3) transformation of a qutrit can be realized with similar resources --- two fields and three steps --- as the general SU(2) transformation of a qubit.
Besides the immediate exponential increase of the Hilbert space qutrits offer other advantages over qubits:
 more secure and efficient quantum communications \cite{cryptography},
 new types of quantum protocols \cite{new protocols,coin tossing},
 new kinds of entanglement \cite{qutrit entanglement},
 larger violations of nonlocality \cite{Bell qutrits}, etc.
To this end, a \qunit~generalization of the Deutsch-Jozsa algorithm has been proposed \cite{Fan},
 which besides its original ability to distinguish between constant and balanced functions, offers some new functionalities.

Grover's quantum search with qudits has also been proposed, where the Hadamard gate, used in the original qubit version, is replaced by a discrete Fourier transform (DFT) \cite{Fan2} or another $\d$-dimensional transformation \cite{Li} in order to construct the \operator~ operator (also known as the diffusion operator). A ternary Grover search was discussed in \cite{WangPerkowski}. We note, however, that these proposals are far from being optimal for a number of reasons, originating mainly from the numerous redundant physical interactions they require, which pose unnecessary challenges to a quantum computer.

In this paper, we introduce a different implementation of multivalued Grover's search, which represents a considerable simplification over the earlier methods \cite{Fan2,Li,WangPerkowski}.
Our implementation of the \operator~operator for a qudit with $\d$ states requires the minimum possible number of physical steps -- just two steps, compared to $2\d$ steps in the proposal of Li \emph{et al.} \cite{Li} and $2\d^2$ steps in the proposal of Fan in \cite{Fan2}.
Moreover, in our implementation of the \operator~ operator, no specific phase relations are required, which makes it far easier to implement than in the two earlier proposals, which impose strict phase relations.
Furthermore, our implementation allows one to use resonant external fields, unlike the implementation proposed by Li \emph{et al.} \cite{Li}, which demands far detuned fields;
resonant interactions allow the fastest implementation because they allow to use the minimum pulse area.
We can therefore claim that our implementation has a double speed-up over the existing proposals: the construction of the \operator~ operator is faster by a factor of $\d$ or $\d^2$,
 and the coupling fields can be on resonance, which is a speed-up by at least a factor of 10 as compared to far-off-resonance fields.
Finally, unlike earlier proposals, our implementation is adapted to deterministic Grover search.

%%%%%%%%%%%%%%%%%%%%%%%%%%%%%%%%%%%%%%%%%%%%%%%%%%%%%%%%%%%%%%%%%%%%%%%%%%%%%%%%%%%%%%%%%%%%%%%%%%%%%%%%%%%%%%%%%%%%%%%%%%
\section{Quantum search with qubits}

%%%%%%%%%%%%%%%%%%%%%%%%%%%%%%%%%%%%%%%%%%%%%%%%%%%%%%%%%%%%%%%%%%%%%%%%%%%%%%%%%%%%%%%%%%%%%%%%%%%%%%%%%%%%%%%%%%%%%%%%%%
\subsection{Overview of Grover search}

Grover's algorithm provides a method for solving the unstructured search problem, which can be stated as follows: given a collection of database
elements $x=1,2,\ldots,\ndata$, and an \emph{oracle function} $f(x)$ that acts differently on one \emph{marked} element $\marked$ to all others,
\begin{equation}
f(x) =\left\{\begin{array}{ll}
1,& x=\marked, \\
0,& x\neq \marked,
\end{array}\right.
\label{oracle_fun}
\end{equation}
find the marked element in as few calls to $f(x)$ as possible \cite{Grover}.

The database is encoded into a set of quantum states; each element is assigned to a corresponding state.
Therefore, each possible search outcome is represented as a basis vector $\ket{x}$ in an $\ndata$-dimensional Hilbert space;
correspondingly, the marked element is encoded by a \emph{marked state} $\ket{\marked}$. Thus one can apply unitary operations
(involving the oracle function) to \emph{superpositions} of the different search outcomes, which are thereby searched in parallel.
The Grover algorithm amplifies the amplitude of the marked state $\ket{\marked}$
using constructive interference, while attenuating all other amplitudes, and locates the marked element in $O(\sqrt{\ndata})$ steps.

Before the execution of the algorithm, the quantum register is prepared in an \emph{equal} superposition of all basis elements \cite{Grover},
\be\label{W_register}
\left\vert \aver\right\rangle =\frac{1}{\sqrt{\ndata}}\sum_{x=1}^{\ndata}\left\vert x\right\rangle,
\ee
with $\ndata=2^n$, where $n$ is the number of qubits.
The algorithm consists of the repeated execution of two sequential operations.

(1) \emph{Oracle query.} The oracle marks the marked state $\ket{\marked}$ in each iteration by shifting its phase, leaving other states unaffected:
 $\M{\marked}{\varphi_\marked}\ket{\marked} = e^{\ii \varphi_{\marked}}\ket{\marked}$.
%In fact, this is a conditional phase gate, which can be expressed as a generalized reflection in Hilbert space:
In fact, this is a conditional phase gate, which is implemented by a generalized reflection in Hilbert space:
\be
\M{\marked}{\varphi_\marked} = \mathbf{1}+ (e^{i\varphi_\marked}-1) \ket{\marked} \bra{\marked}.
\label{Householder}
\ee

(2) \emph{\Operator.} This operation is a reflection about the vector $\ket{\aver}$ with a phase $\varphi_\aver$:
\be\label{reflection}
\M{\aver}{\varphi_\aver} = \mathbf{1}+ (e^{i\varphi_\aver}-1) \ket{\aver} \bra{\aver}.
\ee
It can be constructed with the following operations.

(i) Apply the Hadamard gate,
\be\label{Hadamard}
\H = \tfrac{1}{\sqrt{2}} \left[\begin{array}{rr} 1 & 1 \\ 1 & -1 \end{array}\right],
\ee
to each qubit. This is a single-qubit operation, which can be carried out to all qubits simultaneously.

(ii) Apply a conditional phase shift $\M{0}{\varphi_a}$, with $\ket{0}=\ket{0_1 0_2\cdots 0_n}$, wherein all qubits are in logical state $\ket{0}$.

(iii) Repeat step (i).

It can easily be verified that
\be
\H^{\otimes \nqubits}\M{0}{\varphi_\aver}\H^{\otimes \nqubits} = \M{\aver}{\varphi_\aver}.
\ee

The combined action of the oracle and the \operator~gives the \emph{Grover operator} $\G$,
\be\label{G}
\G = \M{\aver}{\varphi_\aver}\M{\marked}{\varphi_\marked}.
\ee

With the initial state given in Eq.~\eqref{W_register}, and during successive applications of the operator $\G$,
the state vector for the system begins and remains in the two-dimensional subspace defined by the non-orthogonal states $\ket{\marked}$ and $\ket{\aver}$.
Each application of $\G$ amplifies the marked state population until it reaches a maximum value close to unity after $\ng$ iterations,
 at which point the search result can be read out. % by a measurement in the computational basis.

%%%%%%%%%%%%%%%%%%%%%%%%%%%%%%%%%%%%%%%%%%%%%%%%%%%%%%%%%%%%%%%%%%%%%%%%%%%%%%%%%%%%%%%%%%%%%%%%%%%%%%%%%%%%%%%%%%%%%%%%%%
\subsection{Deterministic quantum search}

The problem of how to optimize the quantum search routine by allowing arbitrary $\varphi_\aver$ and $\varphi_\marked$ has been studied extensively \cite{Long1,ketaici,Hoyer}.
It is known that the maximum possible amplitude amplification per step of the marked state arises when the phases $\varphi_\aver$ and $\varphi_\marked$ are both set to $\pi$, as in Grover's original proposal \cite{Grover}.
In this case, however, one does not obtain a unit fidelity in the end.
For a \emph{deterministic} search (unit fidelity) both phases $\varphi_\aver$ and $\varphi_\marked$ must be equal, $\varphi_\aver=\varphi_\marked=\varphi$, where
\be
\varphi = 2 \arcsin \frac{\sin\dfrac{\pi}{4j+2}}{\sin\beta}.
\ee
Here $j=\lfloor\pi/(4\beta)+1/2\rfloor$, $\beta=\arcsin\ndata^{-\frac12}$ and $\lfloor x \rfloor$ denotes the integer part of $x$.
The corresponding minimum number of search steps is \cite{Long1}
\be
\ng = j \text{ or } j+1
\label{NG-Long}
\ee
if, respectively, $(2j+1)\beta$ or $(2j-1)\beta$ is closer to $\pi/2$.
This choice of phases is not unique.
For large $\ndata$, as long as the \emph{phase matching} condition $\varphi_\aver=\varphi_\marked=\varphi$ is satisfied \cite{ketaici},
a high fidelity search can be performed for any value of $\varphi$ in the range $\pi/2\lesssim \varphi \leqslant\pi$
and for certain values of $\varphi$, a deterministic quantum search is possible \cite{Long1}.

%%%%%%%%%%%%%%%%%%%%%%%%%%%%%%%%%%%%%%%%%%%%%%%%%%%%%%%%%%%%%%%%%%%%%%%%%%%%%%%%%%%%%%%%%%%%%%%%%%%%%%%%%%%%%%%%%%%%%%%%%%
\section{Quantum search with \qunits}
%%%%%%%%%%%%%%%%%%%%%%%%%%%%%%%%%%%%%%%%%%%%%%%%%%%%%%%%%%%%%%%%%%%%%%%%%%%%%%%%%%%%%%%%%%%%%%%%%%%%%%%%%%%%%%%%%%%%%%%%%%

%%%%%%%%%%%%%%%%%%%%%%%%%%%%%%%%%%%%%%%%%%%%%%%%%%%%%%%%%%%%%%%%%%%%%%%%%%%%%%%%%%%%%%%%%%%%%%%%%%%%%%%%%%%%%%%%%%%%%%%%%%
\subsection{Generalization of Grover's algorithm}

As in the original implementation with qubits, the implementation that we propose here with \qunits\ begins with the system initialized in an equal-weight superposition $\ket{\ini}$ of all basis states, similar to Eq.~\eqref{W_register} but with arbitrary relative phases.
To do that, first all \qunits\ must reside in the logical states $\ket{0_k}$ $(k=1,2,\ldots,\nqubits)$, the collective state thereby being $\ket{0}=\ket{0_1 0_2\cdots 0_\nqubits}$.
The superposition $\ket{\ini}$ is obtained by applying the same transformation $\F$ independently on all \qunits.
The transformation $\F$ generalizes the Hadamard gate $\H$ used in the original qubit implementation.
It can be achieved by means of any physical interaction, which drives the single-\qunit\ state $\ket{0_k}$ into an equal-weight superposition state,
\be
\F\ket{0_k} = \sum_{q=0}^{\d-1}\xi_q\ket{q_k},
\ee
with $\left|\xi_q\right| = \d^{-1/2}$, in all \qunits\ ($k=1,2,\ldots,\nqubits$).
Thus $\F$ is a representative of a large class of $\qsize$-dimensional unitary matrices, in which the first column contains elements of equal moduli \cite{remark}.
Upon application of $\F$ the collective state is an equal-weight superposition,
\be
\ket{\ini} = \F^{\otimes \nqubits} \ket{0} = \sum_{x=1}^{\ndata}\alpha_x \ket{x},
\ee
wherein $\left|\alpha_x\right|=\ndata^{-1/2}$, with $\ndata=\d^\nqubits$ being the database size, and $\ket{x}$ designates the collective states of $\nqubits$ \qunits.

An example of $\F$ is the discrete Fourier transform $\mathcal{F}$ (DFT), wherein all elements, $\mathcal{F}_{jk}=e^{i\pi jk/\qsize}/\sqrt{\qsize}$, differ only in phase.
The Hadamard gate \eqref{Hadamard} of a qubit is in fact the two-dimensional manifestation of DFT.
However, it is not necessary to assume that $\F$ is indeed the DFT $\mathcal{F}$ because the construction of $\mathcal{F}$ for $\d>2$ may be very demanding;
it requires the synthesis of an entire DFT matrix, which needlessly attempts to transform the whole basis of single
qudit states into a new set of superposition states.
In fact, all we need for the matrix $\F$ is an interaction that creates an equal-weight coherent superposition of the $\d$ states of each \qudit\ starting from the \qudit\ state $\ket{0_k}$ only.
Moreover, the relative phases in this superposition can be arbitrary while there are strict relations for them in the DFT $\mathcal{F}$;
 it is only necessary to use the same $\F$ in all steps.
Of course, the matrix $\F$ must be unitary, i.e. $\F^\dagger=\F^{-1}$, because the interaction is supposed to be coherent.
In general, there are numerous suitable choices for $\F$, of which the respective DFT $\mathcal{F}$ is just one possibility but certainly not the most convenient
one for the reasons given above: the only requirement is that $\F$ is unitary and has a column of elements of equal moduli.
It was a circumvention of DFT that enabled to factor the number 21 in the experiment, described in \cite{NMR21}.

Next one applies repeatedly the Grover operator, which has the same form \eqref{G} as for qubits.
The only difference from Eq.~\eqref{G} is that the Hadamard gate $\H$ is replaced by $\F$, which is contained in the \operator,
\be
\M{\aver}{\varphi} = \F^{\otimes \nqubits}\M{0}{\varphi}(\F^{\dagger})^{\otimes \nqubits}.
\label{GroverOperator2}
\ee
This is a reflection with respect to a state that is an equal superposition of all $\ndata=\d^\nqubits$ possible collective states of a system of
$\nqubits$ \qunits.

The conditional phase shifts $\M{0}{\varphi}$ and $\M{\marked}{\varphi}$ are implemented in the same way as for qubits.
There is a variety of techniques for construction of these gates; for example, efficient methods exist for trapped ions \cite{Ivanov-Grover-D1,Linington,Ivanov-Grover-DN}.

%%%%%%%%%%%%%%%%%%%%%%%%%%%%%%%%%%%%%%%%%%%%%%%%%%%%%%%%%%%%%%%%%%%%%%%%%%%%%%%%%%%%%%%%%%%%%%%%%%%%%%%%%%%%%%%%%%%%%%%%%%
\subsection{Construction of $\F$}

%===================================================================================
\begin{figure}[t]
    \includegraphics[width=0.7\columnwidth]{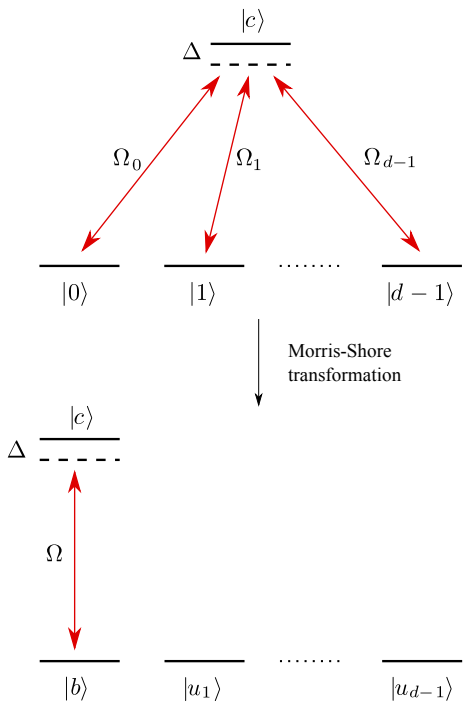}
\caption{(Color online) (Top) \qunit\ in a multipod linkage:  linkage patterns in the original basis (top) and in the Morris-Shore basis (bottom).
The \qunit\ is formed of states $\ket{0}$, $\ket{1}$, $\ldots$, $\ket{\d-1}$, coupled to each other by two-photon Raman processes
 via a common (ancilla) state $\ket{c}$ with a common detuning $\Delta$, but different single-photon Rabi frequencies $\Omega_k$.
In the Morris-Shore basis the multipod reduces to a two-state quantum system formed of a bright state $\ket{b}$ and the original ancilla state $\ket{\d}$, coupled by the \rms\ Rabi frequency $\Omega$.
State $\ket{b}$ is a superposition of the \qunit\ states weighted by the couplings $\Omega_k$; $\ket{\dark_n}$ are uncoupled (dark) states, which do not participate in the dynamics.}
  \label{Npod}
\end{figure}
%===================================================================================

There are potentially numerous ways to implement the interaction $\F$ in different physical systems.
Recently, several different scenarios for the synthesis of an arbitrary SU(3) transformation of a qutrit, including DFT, have been proposed \cite{Vitanov-qutrit}.
Here we will show how the transformation $\F$ can be constructed relatively easily in a multipod system, which is one of the most natural and simplest realizations of \qunits.

A multipod is a system of $\d$ degenerate quantum states $\ket{0}$, $\ket{1}$, $\ldots$, $\ket{\d-1}$, coupled to each other by two-photon Raman processes via a common (ancilla) state $\ket{c}$, as illustrated in Fig.~\ref{Npod} (top).
Such a coupling scheme usually arises in systems of ions or atoms.
$\Omega_k=\left|\Omega_k\right|e^{i\theta_k}$ and $\Delta$ are respectively single-photon (complex) Rabi frequencies and detuning.
If all coupling fields coincide in time, which we assume hereafter, the dynamics of the multipod system is reducible by the Morris-Shore transformation \cite{MShore} to a two-state system,
 as illustrated in Fig. \ref{Npod} (bottom).
The two states are coupled by the root-mean-square (\rms) Rabi frequency $\Omega(t)=\sqrt{\sum_{k=0}^{d-1}\Omega_k^2(t)}$.
Thus the dynamics of the multipod can be derived from the two-state solution; the derivation can be found elsewhere \cite{Kyoseva}.

The propagator for the \qunit\ manifold for \rms\ pulse area $A=\Omega\int_{-\infty}^\infty f(t)dt=2(2l+1)\pi$ (with $l=0,1,2,\ldots$) is given by the generalized reflection $\M{\chi}{\phi}$, with
\be
\ket{\chi}=\frac{1}{\Omega}\left(\Omega_0,\Omega_1,\ldots,\Omega_{\d-1}\right)^T.
\ee
The acquired phase $\phi$ depends on the pulse shape $f(t)$ and the detuning.
For a hyperbolic-secant pulse, $f(t)=\sech(t/T)$, with \rms\ area $A = 2\pi$,
$\phi$ is \cite{Kyoseva}
\be
\phi=\pi-2\arctan (\Delta T).
\ee
The generalized reflection can be created also for other pulse shapes, e.g. Gaussian, but the required pulse area and detuning have to be evaluated numerically.

%===================================================================================
\begin{figure}[t!]
    \includegraphics[width=0.9\columnwidth]{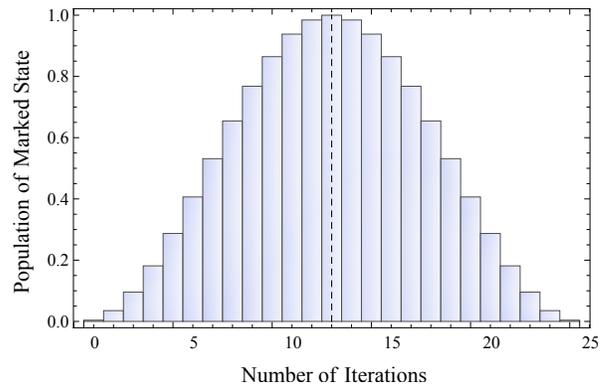}
\caption{(Color online) Simulation of Grover's search in a system of 5 qutrits. The figure depicts the population of the marked state $\ket{\marked}$ vs the number of applications of the Grover operator \eqref{G}.
The vertical dash indicates the time, when unit fidelity is obtained, corresponding to $N_G=12$ search steps, as predicted by Eq. \eqref{NG-Long}.}
  \label{fidelity}
\end{figure}
%===================================================================================

We note that one of the possible implementations of $\F$ is the reflection $\M{\xi}{\pi}$, with $\ket{\xi}=\alpha\left(\sum_{q=0}^{\qsize-1}\xi_q\ket{q_k}-\ket{0_k}\right)$ and $\left|\alpha\right|=1/\sqrt{2(1-\text{Re}\xi_0)}$.
It can be obtained in a multipod system with resonant interaction ($\Delta=0$), wherein the couplings fulfill the following conditions \cite{Kyoseva}
\be
\Omega_0=\sqrt{\frac{1}{2}\left(1-\frac{1}{\sqrt{\qsize}}\right)},\quad
\Omega_{k\neq 0}=\sqrt{\frac{1}{2(\qsize-\sqrt{\qsize})}}.
\ee

%%%%%%%%%%%%%%%%%%%%%%%%%%%%%%%%%%%%%%%%%%%%%%%%%%%%%%%%%%%%%%%%%%%%%%%%%%%%%%%%%%%%%%%%%%%%%%%%%%%%%%%%%%%%%%%%%%%%%%%%%%
\subsection{Example}

Simulation of quantum search  in a system of 5 qutrits ($\d=3$, $\nqubits=5$) is shown in Fig. \ref{fidelity},
 where the probability to find the marked state is plotted as a function of time.
The corresponding database contains $\ndata=\d^\nqubits=243$ elements.
Unit probability is obtained in $N_G = 12$ interaction steps, denoted with a vertical dash, after which it decreases as a part of oscillations between zero and unity in a long run.

%%%%%%%%%%%%%%%%%%%%%%%%%%%%%%%%%%%%%%%%%%%%%%%%%%%%%%%%%%%%%%%%%%%%%%%%%%%%%%%%%%%%%%%%%%%%%%%%%%%%%%%%%%%%%%%%%%%%%%%%%%
\section{Conclusion}
%%%%%%%%%%%%%%%%%%%%%%%%%%%%%%%%%%%%%%%%%%%%%%%%%%%%%%%%%%%%%%%%%%%%%%%%%%%%%%%%%%%%%%%%%%%%%%%%%%%%%%%%%%%%%%%%%%%%%%%%%%

Earlier proposals for Grover's quantum search with qudits use the discrete Fourier transform \cite{Fan2} or other compound transformations \cite{Li}, to assemble the \operator~ operator. These transformations, carried out twice at each search step, demand an increasing number of redundant physical interactions and thus pose unnecessary challenges to a quantum computer. Instead, in our simplified scheme for qudit Grover search we propose to use a reflection operator $\F$, achieved in a \emph{single physical interaction}, which does not even assume any phase relations between the fields driving individual qudits. Our method minimizes the number of algorithmic steps, as well as the duration of each step, which results in a minimal number of interaction steps, fast implementation, increased immunity against detrimental effects of decoherence or inevitable imperfections, resulting from coherent interactions, etc., and deterministic search. Because of its conceptual simplicity, our method is applicable in numerous physical systems. We have shown how $\F$ can be constructed relatively easily, in a single interaction step, in a multipod system, which is one of the most natural and simplest realizations of \qunits.

\acknowledgements
This work is supported by the EU 7th Framework Programme collaborative projects iQIT and FASTQUAST, and the Bulgarian NSF grants D002-90/08 and DMU-03/103.

%%%%%%%%%%%%%%%%%%%%%%%%%%%%%%%%%%%%%%%%%%%%%%%%%%%%%%%%%%%%%%%%%%%%%%%%%%%%%%%%%%%%%%%%%%%%%%%%%%%%%%%%%%%%%%%%%%%%%%%%%%
%%%%%%%%%%%%%%%%%%%%%%%%%%%%%%%%%%%%%%%%%%%%%%%%%%%%%%%%%%%%%%%%%%%%%%%%%%%%%%%%%%%%%%%%%%%%%%%%%%%%%%%%%%%%%%%%%%%%%%%%%%
%%%%%%%%%%%%%%%%%%%%%%%%%%%%%%%%%%%%%%%%%%%%%%%%%%%%%%%%%%%%%%%%%%%%%%%%%%%%%%%%%%%%%%%%%%%%%%%%%%%%%%%%%%%%%%%%%%%%%%%%%%


\begin{thebibliography}{99}
\bibitem{remark} The number of search steps depends on the initial amplitude of the marked state and thereby it may be different for different marked states. Hence, a slight deviation of these moduli from being equal is acceptable, as long as the smallest and the largest of them produce the same number $N_G$.
\end{thebibliography}
\end{document}